\documentclass[%
 reprint,
amsmath,amssymb,
pra,
]{revtex4-2}

\usepackage{xypic}
\usepackage{graphicx}
\usepackage{dcolumn}
\usepackage{bm}
\usepackage{color}

\newcommand{\al}{\alpha}

\newcommand{\ga}{\gamma}

\newcommand{\de}{\delta}

\newcommand{\om}{\omega}
\newcommand{\Om}{\Omega}

\newcommand{\mm}{{\mathcal M}}

\newcommand{\eq}{\equiv}
\newcommand{\ift}{\infty}
\newcommand{\fr}{\frac}

\newcommand{\ra}{\rangle}
\newcommand{\la}{\langle}
\newcommand{\pa}{\partial}
\newcommand{\raw}{\rightarrow}

\newcommand{\non}{\nonumber}

\newcommand{\hh}{\hat{H}}

\newcommand{\hP}{\hat{P}}
\newcommand{\hl}{\hat{L}}

\newcommand{\hb}{\hat{b}}

\newcommand{\hrr}{\hat{\rho}_r}

\newcommand{\ho}{\hat{O}}

\newcommand{\hj}{\hat{J}}

\newcommand{\lo}{\bar{O}}

\newcommand{\rs}{{\rm S}}
\newcommand{\ri}{{\rm I}}

\newcommand{\rt}{{\rm tot}}
\newcommand{\rin}{{\rm int}}
\newcommand{\env}{{\rm E}}
\newcommand{\Tr}{{\rm Tr}}
\newcommand{\rR}{{\rm Re}}
\newcommand{\rI}{{\rm Im}}
\newcommand{\ef}{{\rm eff}}
\newcommand{\rhc}{{\rm H.c.}}

\newcommand{\rap}{{\rm app}}

\newcommand{\bz}{{\bm z}}

\newcommand{\+}{^\dagger}
\newcommand{\bo}{\bar{O}}
\newcommand{\ze}{^{\{0\}}}

\newcommand{\tp}{\tilde{P}}

\newcommand{\3}{\!\!\!}

\newcommand{\dz}{\!\!\!\!\!\!\!\!\!\!\!\!}
\newcommand{\q}{\quad}
\newcommand{\qq}{\quad\quad}

\usepackage{comment}

\begin{document}

\preprint{APS/123-QED}

\title{Positivity-preserving non-Markovian Master Equation for Open Quantum System Dynamics: Stochastic Schr\"odinger Equation Approach}%

\author{Wufu Shi$^{1,3}$}
\email{wufushi2015@hotmail.com}

\author{Yusui Chen$^2$}
\email{yusui.chen@nyit.edu}

\author{Quanzhen Ding$^3$}

\author{Jin Wang$^1$}
\email{jin.wang.1@stonybrook.edu}

\author{Ting Yu$^3$}
\email{tyu1@stevens.edu}

 \affiliation{$^1$Department of Chemistry, Stony Brook University, Stony Brook, New York 11794, USA}
 \affiliation{$^2$Physics Department, New York Institute of Technology, Old Westbury, NY 11568, USA}
\affiliation{$^3$Physics Department, Stevens Institute of Technology, Hoboken, NJ 07030, USA}

\date{\today}

\begin{abstract}

Positivity preservation is naturally guaranteed in exact non-Markovian master equations for open quantum system dynamics. However, in many approximated non-Markovian master equations, the positivity of the reduced density matrix is not guaranteed. In this paper, we provide a general class of time-local, perturbative and positivity-preserving non-Markovian master equations generated from stochastic Schr\"odinger equations, particularly quantum-state-diffusion equations. Our method has an expanded range of applicability for accommodating a variety of non-Markovian environments. We show the positivity-preserving master equation for a three-level system coupled to a dissipative bosonic environment as a particular example to exemplify our general approach. We illustrate the numerical simulations with an analysis explaining why the previous approximated non-Markovian master equations cannot guarantee positivity. Our work provides a consistent master equation for studying the non-Markovian dynamics in ultrafast quantum processes and strong-coupling systems.

\end{abstract}

\maketitle

\section{Introduction}

A density matrix of a quantum system is positive semidefinite, as its eigenvalues naturally are the probabilities of the associated  eigenstates. For a closed system, the positivity of the density matrix is always preserved in the dynamical equations, e.g., the von Neumann equation. However, no quantum systems can be isolated from the surrounding environment. In the context of the theory of open quantum systems (OQSs), the state of the central quantum system is characterized by the reduced density matrix whose time evolution equation is the master equation (ME) instead \cite{quantum_noise_2002, quantum_noise_2010}. Generally, it is extremely difficult to obtain the exact ME due to the infinite number of degrees of freedom of the environment. Several perturbation strategies have been developed in the past decades to obtain approximated MEs~\cite{Diosi1993,Diosi19931}.  For instance, Lindblad-type~\cite{Lindblad} and Redfield-type~\cite{Redfield} MEs based on the Born-Markov approximation effectively describe the Markovian dynamics of many physical processes \cite{MMEapp1,MMEapp2,MMEapp3}. However, among the two typical Markovian MEs, the former can preserve positivity, while the latter cannot \cite{strunz-hug}. 
It is a dilemma to preserve the positivity of MEs with perturbative methods beyond the original Lindblad MEs.

Moreover, the theory of non-Markovian OQSs recently attracted great interests because Markovian approximations are not valid in certain ultrafast processes~\cite{fast1,fast2,fast3,odonnell_high_2020,metzger_harnessing_2017, link_non-markovian_2022}. A comprised solution is to use various \emph{weaker} approximations to maintain certain non-Markovian features beyond the Lindblad ME. Usually, such changes would lead to the new ME which cannot guarantee positivity preservation. Another feasible solution is the hierarchical equations of motion (HEOM) technique~\cite{tanimura,tanimura_numerically_2020}, which is a numerically exact approach to studying the evolution of a density matrix without the typical assumptions that conventional Lindblad or Redfield MEs use. HEOM technique is applicable in computing expectation values of quantum observables, at both zero and finite temperature. But HEOM is not a conventional ME, which is supposed to be a homogeneous equation of the reduced density matrix only. In studying the detailed balance breaking in open quantum systems~\cite{balance2}, the numerically generated density matrix, in chronological order, $\hrr(t)$ is often insufficient to compute the probability flow and analyze the flow's detailed components, while a conventional ME can interpret transition processes between arbitrary two states explicitly. As the result, it is crucial and necessary to obtain a self-consistent ME. And the consequent challenge is twofold: (1) obtaining exact non-Markovian MEs is extremely difficult because of the lack of mathematical tools~\cite{NMME1,NMME2,chen_exact_2014,NMME4,chen_non-markovian_2018,NMME3,chen_exact_2020,NMME5,carballeira_stochastic_2021,Ding_1,Ding_2,Ding_3,qiao_quantumness_2020,Ding2021}; (2) positivity preservation is not guaranteed in perturbative MEs when certain approximations are applied~\cite{vacchini_generalized_2016}. The purpose of this paper is to solve this long-standing problem.

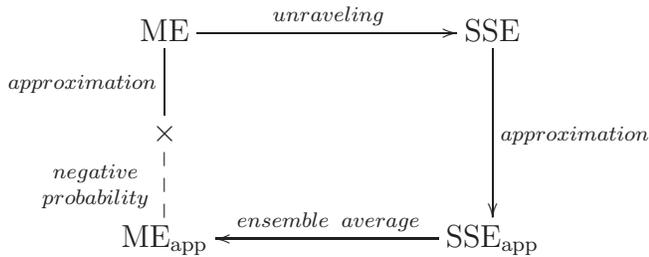
\begin{figure}[h]
\centerline{%
\begin{large}
\xymatrix{
 \rm{ME} \ar[rrr]^{unraveling}\ar@{-}[d]_{approximation} & &  & \rm{SSE} \ar@{->}[dd]^{  approximation} \\ 
  \times \ar@{--}[d]_{\mbox{\footnotesize$\begin{array}{c}
  negative \\
  probability \end{array}$}}    & & &  \\
  \rm{ME_\rap}&       &          & \rm{SSE_\rap} \ar[lll]_{ensemble\;\; average} }  
\end{large}
}
	\caption{The mind map of obtaining positivity-preserving MEs.}\label{rect}
\end{figure}

We demonstrate our solution in Fig.~\ref{rect}. Due to the failure to guarantee positivity, the path of obtaining the consistent approximated ME from the exact ME is blocked. However, in contrast to the dynamics of mixed states of OQSs, the positivity preservation is always guaranteed in the pure state dynamics, even when approximations are applied ($|\psi\ra\la\psi|$ is always positive-semidefinite, here the state $|\psi\ra$ does not have to be normalized). Additionally, the stochastic Schr\"odinger equation (SSE) approach has provided a rigorous method to obtain the associated ME by averaging an infinite number of stochastic pure-state trajectories in the Markov limit~\cite{SSE_Moler93,SSE_Dalibard92,SSE_carmichael93,SSE_Dum92,link_stochastic_2017,jump1,jump2}. 
Thus, we propose a strategy to generate positivity-preserving non-Markovian MEs: (1) start with a formal exact non-Markovian SSE; (2) apply truncation and obtain the approximated SSE; (3) recover the approximated ME from the approximated SSE rigorously.  
 In the end, we assure that the generated approximated ME using the strategy can guarantee positivity as expected.

 If we restrict the measurement on the environment to Bargmann coherent states, the evolution of the stochastic pure state can be characterized by the quantum-state-diffusion (QSD) equation~\cite{QSD_Gisin84,QSD_Gisin92,QSD_Strunz96,QSD_Stunz99,QSD_Yu99,fleming_non-markovian_2012,luo_dynamical_2015}, a time-evolution equation of a stochastic quantum trajectory $|\psi_z\ra \eq \la z||\Psi_{\rt}\ra$, where $||z\ra$ is the Bargmann coherent state of the environment and $z$ represents a large number of complex Gaussian random variables.
 Consequently, the reduced density matrix can be recovered using $\hrr = \Tr_\env[|\Psi_{\rt}\ra\la\Psi_{\rt}|]$, which is equivalent to the ensemble average over all stochastic quantum trajectories $\hrr = \mm(|\psi_z\ra \la \psi_z|)$. As shown in Fig.\ref{rect}, the exact ME and SSE are rigorously equivalent. When some approximations have to be employed, the approximated SSE ($\rm{SSE_\rap}$) can numerically generate the positivity-preserving reduced density matrix $\hrr(t)$ at any time. But the approximated ME, if derived in the same manner, is not positivity-preserving guaranteed. In this paper, we develop a method to generate an approximated ME ($\rm{ME_\rap}$), which is positivity-preserving and has the same reduced density matrix as the ones numerically recovered from the corresponding approximated SSE at any time. 

This paper is organized as follows. In Sec. II, we briefly review the QSD approach and introduce our method to derive the positivity-preserving MEs. In Sec. III, we study a dissipative three-level system and demonstrate how to derive the positivity-preserving ME explicitly. We close with a conclusion in Sec. IV.

\section{General Methods}\label{sec:2}
The theory of OQSs studies the dynamics of a quantum system coupled with an external quantum system or an environment. Generally, the system's dynamics are significantly influenced by the environment, e.g., the quantum decoherence process and the quantum entanglement regeneration process. The total Hamiltonian of the combined system and environment is usually written as:
\begin{equation}
    \hh_{\rt} = \hh_\rs + \hh_{\rin} + \hh_\env.
\end{equation}
Here, the environment Hamiltonian $\hh_\env$ contains an infinite number of bosonic modes. And the interaction Hamiltonian of the coupling between system and environment can be assumed and formally written as:

\begin{eqnarray}
    \hh_\env &=& \sum_k \omega_k \hb_k\+ \hb_k, \non\\ 
    \hh_{\rin} &=& \hl\sum_k g_k \hb_k\+ + \rhc, \non
\end{eqnarray}
where $\hl$ is the system's operator linearly coupled to the environment. We assume that the environment is at zero temperature and the initial state of the combined system and environment is a product state, $|\Psi_{\rt}(t=0)\ra = |\psi_\rs(t=0)\ra\otimes |0_\env\ra$. As mentioned above, having restricted the measurement on the environment to Bargmann coherent states will lead to the quantum trajectory in the form of $|\psi_z\ra \eq \la z||\Psi_{\rt}\ra$, where $||z\ra \eq \otimes_k ||z_k\ra$ is the Bargmann coherent state of the entire environment, satisfying $\hb_k ||z\ra = z_k||z\ra$. The evolution of $|\psi_z\ra$ is governed by the QSD equation.

\subsection{Quantum-state-diffusion approach}

In the environmental interaction picture, the interaction Hamiltonian reads:
\begin{eqnarray}
\hh_{\rin}^\ri(t)= \hl \sum_k g_k \hb_k\+ e^{i\omega_k t} + \hl\+ \sum_k g_k^* \hb_k e^{-i\omega_k t}.
\end{eqnarray}
Using the identity resolution of Bargmann coherent states
\begin{eqnarray}
\hat{I}_\env  = \int \frac{d^2 z}{\pi} e^{-|z|^2} ||z\ra \la z||,
\end{eqnarray}
the Schr\"{o}dinger equation, regarding the evolution of the pure state of the composite system $|\Psi_{\rt}\ra$, can be rewritten in the Bargmann space representation (setting $\hbar =1$),
\begin{eqnarray}
\partial_t \la z||\Psi_{\rt}\ra &=& -i \la z|| (\hh_\rs +\hh_{\rin}^\ri(t)) |\Psi_{\rt}\ra \non \\
&=& -i(\hh_\rs + \hl \sum_k g_k z_k^* e^{i\omega_k t}  )\la z||\Psi_{\rt}\ra  \non \\
& & -i  \hl\+ \sum_k g_k^* e^{-i\omega_k t}\frac{\partial}{\partial z_k^*}\la z||\Psi_{\rt}\ra . \label{4}
\end{eqnarray}
Here, we define a complex Gaussian process $z_t^* \eq -i\sum_k g_k z_k^* e^{i\omega_k t}$, which satisfies the following relations: $\mm(z_t) = \mm(z_t z_s) =0$, and $\alpha(t,s) \eq \mm(z_tz_s^*) =\sum_k |g_k|^2e^{-i\omega_k(t-s)}$, where $\alpha(t,s)$ is the correlation function of the complex Gaussian process $z_t^*$. Then, using the chain rule $\frac{\partial (\cdot) }{\partial z_k^*} = \int_0^t ds \frac{\partial z_s^*}{\partial z_k^*} \frac{\delta (\cdot)}{\delta z_s^*}$, Eq.~(\ref{4}) can lead to a formal linear non-Markovian QSD equation:
\begin{eqnarray}
\partial_t|\psi_z\ra = (-i\hh_\rs+z_t^*\hl - \hl\+\int_0^tds\alpha(t,s)\delta_{z_s^*}) |\psi_z\ra. \label{oqsd}
\end{eqnarray}
By taking the statistical mean over all trajectories, the reduced density matrix of the system can be recovered
 \begin {eqnarray}
  \hrr =\mm(|\psi_z\ra\la\psi_z|).
 \end {eqnarray}
The functional derivative term in Eq.~(\ref{oqsd}) can be formally written using a to-be-determined operator  $\ho$, defined as
\begin {eqnarray}
 \ho(t,s) |\psi_z\ra \eq \de_{z^*_s} |\psi_z\ra,
\end {eqnarray}
which can be solved through an operator evolution equation:
 \begin {eqnarray}
  \pa_t \ho(t,s) =-i[\hh_\ef, \ho(t,s)] -i\de_{z^*_{s}} \hh_\ef , \label{paro}
 \end {eqnarray}
 where $\hh_\ef$ is the effective Hamiltonian:
 \begin{eqnarray}
 \hh_\ef \eq \hh_\rs +iz_t^*\hl -i\hl\+\bar{O},
 \end{eqnarray}
and $\bar{O}(t) \eq \int_0^t ds\, \al(t,s) \ho(t,s)$. Therefore, the formal linear QSD equation reads:
\begin{eqnarray}
 \pa_t |\psi_z\ra = (-i\hh_\rs +z_t^*\hl -\hl\+\bar{O}(t))|\psi_z\ra.
\end{eqnarray}

Generally, the structure of the exact O-operator is complicated. Only a few models can be solved with the exact O-operator. A compromised solution to this difficulty is to replace the exact O-operator with an approximated one. One can drop certain terms of the  O-operator to simplify the calculation, called a truncation operation. How to truncate the O-operator depends on the type of interaction and the size of the system. Without the loss of generality, the O-operator can be written as a sum of all component operators~\cite{JJ-XYZ}:
\begin{eqnarray}
\ho(t,s,z^*) = \sum_n \ho_n(t,s,z^*). 
\end{eqnarray}
And the approximated O-operator after the truncation is defined as:
\begin{eqnarray}
\ho^N(t,s,z^*) \eq \sum_{n=0}^N \ho_n(t,s,z^*). 
\end{eqnarray}
Consequently, the approximated QSD equation after the truncation reads:
\begin{eqnarray}
\partial_t|\psi^N_z\ra = (-i\hh_\rs +z_t^*\hl - \hl\+\int_0^tds\alpha(t,s)\ho^N) |\psi^N_z\ra. \label{eq:13}
\end{eqnarray}
where the trajectory $|\psi_z^N\ra$ is the associated approximated trajectory. 

One of the advantages of the non-Markovian QSD approach is that any reduced density operator  recovered from quantum trajectories $\hrr = \mm(|\psi_z\ra \la\psi_z|)$ is always positivity preserved, even if quantum trajectories are numerically generated by the approximated QSD equation~(\ref{eq:13}), $\hrr^N = \mm(|\psi^N_z\ra \la \psi^N_z|)$. (For the single trajectory, we know that $|\psi_z^N\ra\la\psi_z^N|$ must be positive semidefinite. The ensemble average $\mm(|\psi_z^N\ra\la\psi_z^N|)$ can be considered as a convex combination of $|\psi_z^N\ra\la\psi_z^N|$, therefore, $\mm(|\psi_z^N\ra\la\psi_z^N|)$ is also positive semidefinite. Principally, this is how we derive the positivity-preserving ME from the approximated QSD equation.

\subsection{Positivity-preserving ME}

For a given exact QSD equation, the associated ME reads:
\begin{eqnarray}
 \pa_t \hrr &=& \mm(\frac{\partial |\psi_z\ra}{\partial t} \la \psi_z| + |\psi_z\ra \frac{
\pa \la \psi_z|}{\pa t}) \non\\
 &=& \mm(-iH_\ef |\psi_z\ra\la \psi_z| + i |\psi_z\ra\la \psi_z| H_\ef\+) \non \\
 &=& -i[\hh_\rs, \hrr] +\hl\mm(z_t^*\hP) +\mm(z_t\hP)\hl\+  \non \\
 & & -\hl\+\mm(\bo\hP) -\mm(\hP\bo\+)\hl, \label{ME}
\end{eqnarray}
where $\hP$ denotes the stochastic  operator $\hP \eq |\psi_z\ra\la\psi_z|$. Using the Novikov theorem (see Appendix \ref{Novikov}), the two terms, $\mm(z_t^*\hP)$ and $\mm(z_t\hP)$\textcolor[rgb]{1,0,0}{,} in the above equation\textcolor[rgb]{1,0,0}{,} can be estimated,
\begin{eqnarray}
 \mm(z_t^*\hP) = \int_0^t ds \alpha^*(t,s) \mm(\delta_{z_s} \hP) = \mm(\hP \bo\+). \label{novi}
\end{eqnarray}
As a result, the formal ME is obtained,
\begin{eqnarray}
 \pa_t \hrr =-i[\hh_\rs, \hrr] + [\hl, \mm(\hP \bo\+)]  - [\hl\+, \mm(\bo\hP)]. \label{exactME}
\end{eqnarray}
The above-derived ME is positivity-preserving since the reduced density matrix $\hrr$ is rigorously equivalent to the exact stochastic quantum trajectory governed by the QSD equation (\ref{oqsd}). 

Next, we will demonstrate why the ME can not guarantee positivity if all the four exact  O-operators in Eq.~(\ref{exactME}) are replaced by the approximated one $\ho^N$. Following the similar method of obtaining Eq.~(\ref{exactME}), the approximated ME for the perturbative reduced density matrix $\hrr'$ reads:
\begin{eqnarray}
 \pa_t \hrr' =-i[\hh_\rs, \hrr'] + [\hl, \mm(\hP' (\bo^{N\dagger})]  - [\hl\+, \mm(\bo^N\hP')]. \non \\ \label{appME}
\end{eqnarray}
However, it is worth pointing out that the reduced density matrix $\hrr'$ can violate positivity because the approximated ME can not be unraveled by the QSD equation (\ref{eq:13}). To clarify the difference between Eq.~(\ref{appME}) and the approximated ME which is rigorously equivalent to the approximated QSD equation (\ref{eq:13}), we recover the ME starting from the identity $\hrr^N = \mm(|\psi_z^N\ra\la\psi_z^N|)$ and the QSD equation (\ref{eq:13}). The approximated ME reads
\begin{eqnarray}
\pa_t \hrr^N 
 &=& -i[\hh_\rs, \hrr^N] +\hl\mm(z_t^*\hP^N) +\mm(z_t\hP^N)\hl\+  \non \\
 & & -\hl\+\mm(\bo^N\hP^N) -\mm(\hP^N \bo^N{}\+)\hl, \label{appME2}
\end{eqnarray}
where $\hP^N \eq |\psi_z^N\ra\la\psi_z^N|$.
After applying the Novikov theorem to simplify the term of $\mm(z_t^*\hP^N)$, it is easy to verify that in the general case
\begin{eqnarray}
\mm(z_t^*\hP^N) = \int_0^t ds\, \al^*(t,s)\mm(\de_{z_s}\hP^N) \neq  \mm(\hP^N \bo^{N\dagger}). \non
\end{eqnarray}
This is why the above-mentioned approximated ME (\ref{appME}) cannot be unraveled by the QSD equation (\ref{eq:13}). To solve this problem, we need to know the exact value of $\de_{z_s}\hP^N$, therefore a new O-operator has to be introduced
\begin{eqnarray}
 \ho_d(t,s,z^*) |\psi_z^N\ra \eq \delta_{z_s^*}|\psi_z^N\ra,
\end{eqnarray}
where the new operator $\ho_d(t,s,z^*)$ is determined by an evolution equation,
\begin{eqnarray}
 \pa_t \ho_d(t,s,z^*) &=& [-i\hh_\rs + z_t^*\hl - \hl\+ \bo^N, \ho_d(t,s,z^*)] \non \\
 & & -\hl\+\delta_{z_s^*} \bo^N, \label{eq:20}
\end{eqnarray}
together with the initial condition,
\begin{eqnarray}
 \ho_d(t=s,s,z^*) =\hl.
\end{eqnarray}
The subtle difference between $\ho^N$ and $\ho_d$ is just the reason of positivity violation in the ME (\ref{appME}).
Consequently, the result of applying the Novikov theorem is revised to
\begin{eqnarray}
 \mm(z_t^*\hP^N) &=& \int_0^t ds\, \al^*(t,s)\mm(\de_{z_s}\hP^N) \non \\
 &=& \mm(\hP^N\bo_d\+), \label{eq:22}
\end{eqnarray}
where $\bo_d\+(t,z^*) \eq \int_0^t ds \al^*(t,s)\ho_d\+(t,s,z^*)$. By substituting Eq.~(\ref{eq:22}) into Eq.~(\ref{appME2}), we obtain the formal positivity-preserving approximated ME,
\begin{eqnarray}
\pa_t \hrr^N &=& -i[\hh_\rs,\hrr^N] + \hl \mm(\hP^N \bo_d\+) -  \mm(\hP^N \bo^{N\dagger})\hl \non \\
& & -\hl\+ \mm(\bo^N \hP^N) + \mm(\bo_d \hP^N) \hl\+. \label{appMEfinal}
\end{eqnarray}

\section{Models and Results}

Here, we consider a ladder-type three-level system coupled with a dissipative zero-temperature reservoir and use it to demonstrate how to obtain the positivity-preserving approximated ME. The total Hamiltonian reads
\begin {eqnarray}
 \hh_\rt = \om \hj_z +\sum_k g_k (\hj_+\hb_k +\hb_k\+\hj_-) +\sum_k \om_k \hb_k\+\hb_k,
\end {eqnarray}
where $g_k$ is the real coupling strength of the $k$th mode. $\hj_+$ ($\hj_-$) is the raising (lowering) operator of the three-level system, satisfying the commutation relation $\hj_z = \frac{1}{2}[\hj_+, \hj_-]$. The three operators have the matrix form
\begin{eqnarray}
\hj_z = \begin{bmatrix}
1 & 0 & 0\\
0 & 0 & 0 \\
0 & 0 & -1
\end{bmatrix}, 
\hj_+ = \sqrt{2}\begin{bmatrix}
0 & 1 & 0\\
0 & 0 & 1 \\
0 & 0 & 0
\end{bmatrix}, 
\hj_- = \sqrt{2}\begin{bmatrix}
0 & 0 & 0\\
1 & 0 & 0 \\
0 & 1 & 0
\end{bmatrix}. \non
\end{eqnarray}

\subsection{Master equation for the three-level system}
In Ref.~\cite{Junprl,chen_exact_2020}, it has been proved that the O-operator of a dissipative three-level system contains noise up to the first order. We use a noise-free operator $\ho\ze$ to replace the exact $\ho$ in the QSD equation:
\begin {eqnarray}
 \pa_t |\psi_z\ze\ra =(-i\hh_\rs +z^*_t\hl -\hl\+\lo\ze)|\psi_z\ze\ra, \label{appqsd}
\end {eqnarray}
where $\lo\ze \eq \int_0^t ds \al(t,s) \ho\ze(t,s)$, and the Lindblad operator $\hl =\hj_-$. The operator $\ho\ze$ is governed by its evolution equation
\begin {eqnarray}
 \pa_t \ho\ze(t,s) =[-i\hh_\rs -\hl\+\lo\ze,\ho\ze(t,s)],  \label{6O0eq}
\end {eqnarray}
with its initial condition
\begin{eqnarray}
 \qq \ho\ze(t=s,s)=\hl.
\end{eqnarray}
According to Eq.~(\ref{6O0eq}), the equation is valid only when the operator $\ho\ze$ has the form of
\begin{eqnarray}
 \ho\ze(t,s) \eq f_1(t,s)\hj_- +g_1(t,s)\hj_z\hj_-, \label{ansatz}
\end{eqnarray}
where $f_1(t,s)$ and $g_1(t,s)$ are two to-be-determined evolution coefficients. By substituting the ansatz (\ref{ansatz}) into Eq.~(\ref{6O0eq}), the coefficients $f_1$ and $g_1$ can be determined by the following differential equations
\begin{eqnarray}
  \pa_t f_1 &=& (i\om +2G_1) f_1,  \non \\
  \pa_t g_1 &=& (-2F_1 +4G_1)f_1 +(i\om +2F_1 -2G_1) g_1, \;\; \label{3fh}
\end{eqnarray}
associated with the initial conditions,
\begin{eqnarray}
  f_1(t=s,s)=1, \qq  g_1(t=s,s) =0, 
\end{eqnarray}
where $F_1(t) \eq \int_0^t ds \al(t,s) f_1(t,s)$, and $G_1(t) \eq \int_0^t ds \al(t,s) g_1(t,s)$.
Subsequently, the time-evolution equation of the operator $\ho_d(t,s,z^*) $ reads
\begin{eqnarray}
  \pa_t \ho_d &=& [-i\hh_\rs + z_t^*\hl - \hl\+ \bo\ze, \ho_d]. \label{Odevo}
\end{eqnarray}
Note that the last functional derivative term in Eq.~(\ref{eq:20}) has been eliminated because $\ho\ze$ is noise-free. Similarly, the ansatz of the operator $\ho_d$ reads
\begin{eqnarray}
  \ho_d(t,s,\bz^*) &\eq& f_2(t,s)\hj_- +g_2(t,s)\hj_z\hj_- \non \\
  &&  +\int_0^t ds' p_2(t,s,s') z^*_{s'}\hj_-^2. \label{Odansatz}
\end{eqnarray}
By substituting the ansatz (\ref{Odansatz}) into Eq.~(\ref{Odevo}), the new set of coefficients, $f_2(t,s), g_2(t,s)$ and $p_2(t,s,s')$, are determined by
\begin{eqnarray}
 \pa_t f_2 &=& (i\om +2G_1)f_2,  \non \\
 \pa_t g_2 &=& (-2F_1 +4G_1)f_2 +(i\om +2F_1 -2G_1) g_2,  \;\; \non \\
 \pa_t p_2 &=& (2i\om +2F_1) p_2,  \label{3fn}
\end{eqnarray}
with the initial conditions
\begin{eqnarray}
 \q f_2(t=s,s) &=& 1, \non \\
 \q  g_2(t=s,s)&=& 0, \non \\
 p_2(t=s',s,s') &=&  g_2(s',s). \non
\end{eqnarray}
By comparing Eq.~(\ref{3fh}) and Eq.~(\ref{3fn}), it is clear that $f_1=f_2$ and $g_1=g_2$, since they have the same evolution equations and the same initial conditions. As a result, $f_1$ and $g_1$, in the rest of the paper, will be replaced by $f_2$ and $g_2$, respectively.

After obtaining operators $\ho_d$ and $\ho\ze$, the formal ME (\ref{appMEfinal}) for the dissipative three-level model reads
\begin{eqnarray}
 \dz\3\pa_t \hrr\ze \3&=&\3 -i[\hh_\rs, \hrr\ze] +\{ [(F_2\hj_- +G_2\hj_z\hj_-)\hrr\ze, \hj_+]  \non \\
 \dz\3&&\3 +\hj_-^2\int_0^t ds P_2(t,s)\mm(z_s^*\hP\ze)\hj_+\} +\rhc , \label{3Master2}
\end{eqnarray}
where $F_2(t)$, $G_2(t)$, and $P_2(t,s')$ are defined as
\begin{eqnarray}
  F_2(t) &\eq& \int_0^t ds \al(t,s) f_2(t,s), \non \\
  G_2(t) &\eq& \int_0^t ds \al(t,s) g_2(t,s),  \non \\
  P_2(t,s') &\eq& \int_0^t ds \al(t,s) p_2(t,s,s'). \non
\end{eqnarray}

Applying the Novikov theorem and the termination condition in Ref.~\cite{chen_exact_2014}, the term $\hj_-^2\int_0^t ds P_2(t,s)\mm(z_s^*\hP\ze)\hj_+$ in the above ME can be further simplified to 
\begin{eqnarray}
 & &\hj_-^2\mm(z_s^*\hP\ze)\hj_+ \non \\
 &=& \int_0^t ds' \al^*(s,s') \hj_-^2\mm(\hP\ze\ho_d\+(t,s',\bz))\hj_+  \non \\
 &=& \int_0^t ds' \al^*(s,s') f^*_2(t,s') \hj_-^2 \hrr\ze \hj_+\hj_+.  
\end{eqnarray}
Subsequently, the approximated positivity-preserving ME reads,
\begin{eqnarray}
 \3\pa_t \hrr\ze &=& -i[\hh_\rs, \hrr\ze] +\{ [(F_2\hj_- +G_2\hj_z\hj_-)\hrr\ze, \hj_+] \non\\
 &&  + P_{f^*}\hj_-^2 \hrr\ze\hj_+^2\} +\rhc,
\end{eqnarray}
where the coefficient $P_{f^*} (t) \eq \int_0^t\int_0^t dsds' P_2(t,s) \al^*(s,s') f^*_2(t,s')$. Note that $\al(t,s)$ is the time correlation function, corresponding to a variety of spectra, white or colored. For simplicity, we assume the environment is described by an Ornstein-Uhlenbeck process. So, its correlation function is $\al(t,s) =a\ga e^{-\ga|t-s|}e^{-i\Om(t-s)}$, where $1/\gamma$ is the scale of memory time and $\Omega$ is the central frequency of the environment. As a result, the coefficients' evolution equations can be simplified from integrodifferential equations to differential equations that
\begin{eqnarray}
 \dz \pa_t F_2 \3&=&\3 a\ga +(i\om -\ga -i\Om +2G_2)F_2, \non \\
 \dz \pa_t G_2 \3&=&\3 -2F_2^2 +(i\om -\ga -i\Om +6F_2 -2G_2)G_2,  \non \\
 \dz \pa_t \tp_2 \3&=&\3 a\ga G_2 +(2i\om -2\ga -2i\Om +2F_2)\tp_2, \non\\
 \dz \pa_t P_{f^*} \3&=&\3 (i\om -\ga -i\Om +2F_2 +2G^*_2)P_{f^*} +\tp_2 +G_2 F_2^*, \non\\
\end{eqnarray}
with the initial conditions,
\begin{eqnarray}
  \q F_2(t=0) =G_2(t=0)= \tp_2(t=0) = P_{f^*}(t=0)=0, \non
\end{eqnarray}
where $\tp_2(t) \eq \int_0^t ds \al(t,s) P_2(t,s)$. 

\subsection{Numerical results}

In this section, we compare the simulation results of the population of states of the three-level system using four different methods: (1) the exact ME, $\hrr = \mm(|\psi_z\ra\la\psi_z|)$ in Eq.~(\ref{exactME}); (2) the approximated positivity-preserving ME, $\hrr\ze = \mm(|\psi_z\ze\ra\la\psi_z\ze|)$ in Eq.~(\ref{appMEfinal}); (3) the approximated non-positivity-preserving ME, $\hrr'$ in Eq.~(\ref{appME}); (4) the approximated QSD, $|\psi_z\ze\ra$ in Eq.~(\ref{appqsd}).

\begin{figure}[ptb]
\begin{center}
\includegraphics[trim={0.23in 0.25in 0.0in 0.0in},clip,
width= 3.8in]
{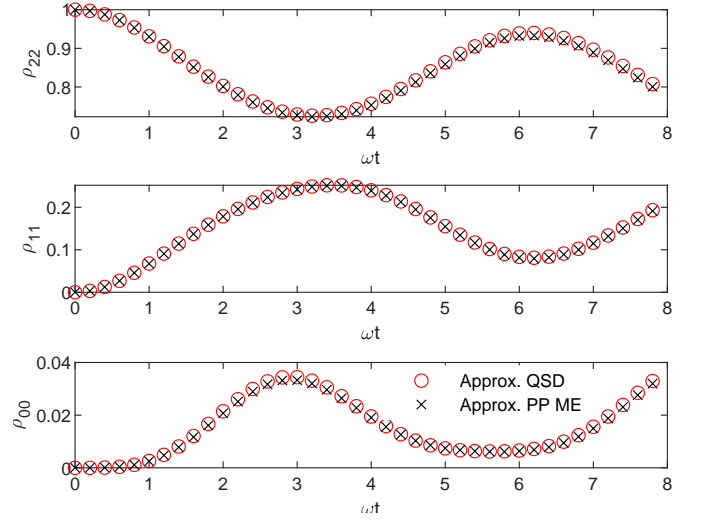}
\end{center}
\caption{Time evolution of the population of the dissipative three-level system, $\rho_{00}$, $\rho_{11}$ and $\rho_{22}$ are generated by two different methods: approximated linear QSD equation approach (red circle), and the derived positivity-preserving approximated ME (black cross). The parameters are $\om=1$, $a =0.8$, $\ga = 0.05$, and $\Om=0$. (The result of the approximated QSD equation approach is obtained by averaging over 5000 quantum trajectories.) }
\label{1}
\end{figure}

First of all, we plot the time evolution of the population of the dissipative three-level system, $\rho_{00}$, $\rho_{11}$, and $\rho_{22}$, generated by approximated QSD equation approach and the approximated positivity-preserving ME approach. The initial state of the system is prepared at an excited state: $|\psi_z(t=0)\ra=|2\ra$. We choose the frequency, $\om =1$, and the central frequency of the environment $\Om = 0$. In a strong non-Markovian regime, $\ga = 0.05$, the simulation results from two methods, as shown in Fig.~\ref{1}, overlap each other. Since the reduced density matrix generated from the approximated QSD approach is naturally positivity-preserving, the matched dynamics prove that our approximated ME gives rise to the same degree of accuracy as the approximated QSD equation. It indicates that the approximated ME can guarantee positivity.

\begin{figure}[ptb]
\begin{center}
\includegraphics
[trim=0in 0in 0in 0in, 
width=3.5in
]{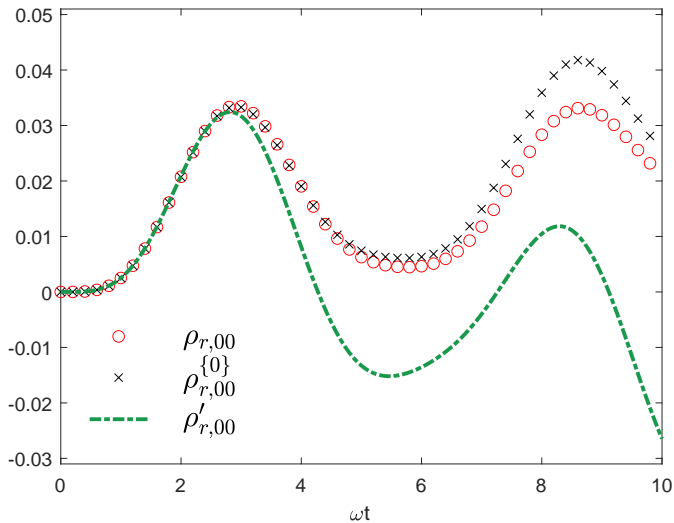}
\end{center}
\caption{ Comparison of the time evolution of the population of the ground state, generated by three different ME approaches: exact ME ($\rho_{r,00}$, red circle), positivity-preserving approximated ME ($\rho\ze_{r,00}$, black cross), and the non-positivity-preserving approximated ME ($\rho'_{r,00}$, green dash-dotted line). The parameters are $\om=1$, $a =0.8$, $\ga = 0.05$, and $\Om=0$. }
\label{2}
\end{figure}

Next, we plot the time evolution of the population of the ground state using three different ME approaches in Fig.~\ref{2}. Using the same parameters as Fig.~\ref{1}, we observe that the exact ME and our approximated positivity-preserving ME both preserve the positivity. Meanwhile, the simulation result of the non-positivity-preserving ME leads to failure due to the appearance of negative probabilities in some time intervals. Furthermore, the magnitude of the negative probability increases with time up to infinity. Consequently, the probabilities of the other two levels also increase to infinity simultaneously. If simply replaces the exact O-operator in the exact ME with the truncated operator $\ho\ze$, then the Eq.~(\ref{appME}) can be explicitly written as
\begin{eqnarray}
\pa_t \hrr' = -i[\hh_\rs, \hrr'] +\big( [(F_2\hj_- +G_2\hj_z\hj_-)\hrr', \hj_+] +\rhc \big).  \non
 \end{eqnarray}
It is clear that the above approximated ME does not preserve positivity in some parameter regions, and may lead to meaningless physics.

\begin{figure}[ptb]
\begin{center}
\includegraphics
[trim=0in 0in 0in 0in, 
width=3.5in
]{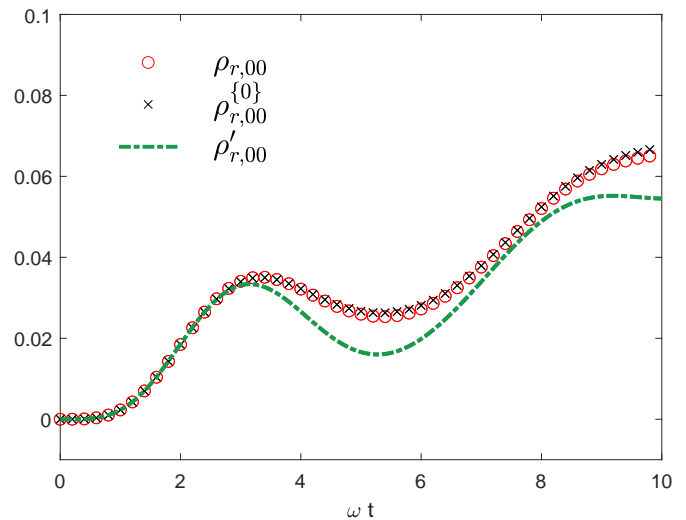}
\end{center}
\caption{ Comparison of the time evolution of the population of the ground state, generated by three different ME approaches: exact ME ($\rho_{r,00}$, red circle), positivity-preserving approximated ME ($\rho\ze_{r,00}$, black cross), and the non-positivity-preserving approximated ME ($\rho'_{r,00}$, green dash-dotted line), in a \emph{moderate} non-Markovian regime. The parameters are $\om=1$, $a =0.2$, $\ga = 0.2$, and $\Om=0$. }
\label{3}
\end{figure}

To further demonstrate the importance of our method in studying non-Markovian dynamics, we plot Fig.~\ref{3}, the time evolution of the population of the ground state in a \emph{moderate} non-Markovian regime. When $\ga = 0.2$, a shorter memory time compared with the parameter $\ga=0.05$ used in Fig.~\ref{2}, the dynamics simulated by the approximated non-positivity-preserving ME $\rho_{r,00}'$ do not contain any negative probabilities. However, its distance from the results of the exact ME approach is significantly larger compared with the results of our positivity-preserving ME. Comparing Figs.~\ref{2} and \ref{3}, we show that the reduced density matrix $\rho_{r}\ze$ can guarantee positivity preservation from the Markovian to the strong non-Markovian regime. In contrast, the reduced density matrix $\rho_{r}'$ cannot offer such confidence. Moreover, for different models and interested parameter spaces, our method is flexible for different approximations. It provides a robust method to obtain positivity-preserving MEs for the analysis of non-Markovian dynamics.

\section{Conclusion}
We have addressed a long-standing issue in OQSs on how to construct positivity-preserving approximated MEs in a general situation. Although several developed MEs, such as Lindblad and Redfield MEs, can provide powerful and efficient mathematical tools, these approaches have common shortcomings since they are rooted in the Born-Markov approximation. In this study, we start from the fact that a reduced density matrix must carry over positivity if recovered from the ensemble average over the stochastic pure states. Then we consider a class of linear approximated QSD equations, exploring the possibility of constructing MEs rigorously equivalent to QSD equations. We find that previous works mistakenly used the approximation relation $\delta_{z_s^*}|\psi_z^N\ra  \approx \ho |\psi_z^N\ra \approx \ho^N|\psi_z^N\ra$. In fact, $\delta_{z_s^*}|\psi_z^N\ra$ should be equal to $\ho_d |\psi_z^N\ra$, where $\ho_d$ is a newly defined operator. No matter how small the difference between the $\ho_d$ and the approximated $\ho^N$ is, replacing $\ho_d$ by $\ho^N$ in the derivation may lead to a violation of positivity of MEs. Consequently, it is necessary to introduce two different approximated $\ho$ to generate the positivity-preserving approximated ME. Generally, we explain why applying an approximation directly on the exact ME may violate positivity, while applying the same approximation on the exact SSE will not. 

In the paper, we propose a general class of positivity-preserving non-Markovian MEs generated from SSEs, in particular, QSD equations.  We explicitly derive the approximated positivity-preserving ME for a dissipative three-level system as a specific example of our general results. Moreover, our simulations also show that the negative probability generated by non-positivity-preserving MEs sometimes ends up with negative infinity, which is definitely not a trivial issue. 

In summary, we have developed a systematic method to obtain a class of approximated but positivity-preserving non-Markovian MEs originating from approximated linear QSD equations. With such MEs, it is feasible to study nonequilibrium dynamics in living or biological systems, perform reliable error analysis for quantum engineering, and investigate dynamics and phase transitions in many-body systems.


\section*{Acknowledgments}
We thank Lajos Di\'{o}si and Water Strunz for useful conversations on the topic over a long period of time.

\appendix



\section{Novikov Theorem} \label{Novikov} 
Firstly, we recall several notations:
\begin{eqnarray}
\mm(\cdot) &=& \int \frac{d^2 z}{\pi} e^{-|z|^2} (\cdot),
\end{eqnarray}
where $\fr{d^2 z}{\pi} \eq \fr{d^2 z_1}{\pi} \fr{d^2 z_2}{\pi} ...\fr{d^2z_k}{\pi} ...$, $|z|^2 \eq \sum_k |z_k|^2$,
$\int d^2z_k \eq \int_{-\ift}^\ift \int_{-\ift}^\ift dx_kdy_k$, and $x_k\eq\rR(z_k)$, $y_k\eq\rI(z_k)$. In addition, the complex Gaussian process is defined as
\begin{eqnarray}
z_t^* \eq -i\sum_k g_kz_k^*e^{i\omega_k t}.
\end{eqnarray}
Consequently, the left-hand side of the Eq.~(\ref{novi}) can be explicitly expanded as
\begin{eqnarray}
\mm(z_t^* \hP) &=& \int \frac{d^2 z}{\pi} e^{-|z|^2}z_t^*\hP \non \\
&=& \int \frac{d^2 z}{\pi} e^{-|z|^2} \big(-i\sum_k g_kz_k^*e^{i\om_kt}  \big)\hP \non\\
&=& \int \frac{d^2 z}{\pi} e^{-|z|^2} \big(-i\sum_k g_k (x_k -iy_k)e^{i\om_kt}  \big)\hP.  \;\;\; \non\\
\label{zt}
\end{eqnarray}
Using integration by parts for the $k$th mode, we have
\begin{eqnarray}
& &\int d^2 z_k e^{-|z_k|^2} x_k\hP \non\\
&=& -\fr{1}{2}\int d^2z_k \fr{\pa}{\pa x_k} (e^{-|z_k|^2}) \hP \non \\
&=& -\fr{1}{2}\int dy_k  ( e^{-|z_k|^2 }\hP ) \big|^{x_k=\ift}_{x_k=-\ift} +\fr{1}{2} \int d^2z_k e^{-|z_k|^2 } \pa_{x_k} \hP. \non \\
\end{eqnarray}
Usually, the boundary terms $(e^{-|z_k|^2 }\hP)|_{x_k \raw \ift}$ and $(e^{-|z_k|^2 }\hP)|_{x_k \raw -\ift}$ in the above row converge to zero rapidly. Thus, we have
\begin{eqnarray}
\int d^2 z_k e^{-|z_k|^2} x_k\hP =\fr{1}{2} \int d^2z_k e^{-|z_k|^2 } \pa_{x_k} \hP. \label{xk}
\end{eqnarray} 
Similarly, we also have
\begin{eqnarray}
\int d^2 z_k e^{-|z_k|^2} y_k\hP =\fr{1}{2} \int d^2z_k e^{-|z_k|^2 } \pa_{y_k} \hP. \label{yk}
\end{eqnarray} 
According to the chain rule, we have
\begin{eqnarray}
    \begin{bmatrix}
        \pa_{x_k} \\
        \pa_{y_k}
    \end{bmatrix} &=&
    \begin{bmatrix}
        \pa z_k /\pa x_k & \pa z_k^*/ \pa x_k \\
        \pa z_k /\pa y_k & \pa z_k^*/ \pa y_k
    \end{bmatrix}
    \begin{bmatrix}
        \pa_{z_k} \\
        \pa_{z_k^*}
    \end{bmatrix} \non\\
    &=&
    \begin{bmatrix}
        1 & 1 \\
        i & -i
    \end{bmatrix}
    \begin{bmatrix}
        \pa_{z_k} \\
        \pa_{z_k^*}
    \end{bmatrix}. \label{eq:a7}
\end{eqnarray}
By employing Eqs.~(\ref{xk}), (\ref{yk}), and (\ref{eq:a7}), we obtain the conclusion:
\begin{eqnarray}
    && \int d^2 z_k e^{-|z_k|^2} z_k^*\hP \non\\
    &=& \int d^2 z_k e^{-|z_k|^2} (x_k -iy_k)\hP \non\\
    &=& \fr{1}{2}\int d^2 z_k e^{-|z_k|^2} 
    \begin{bmatrix}
        1 & -i         
    \end{bmatrix}
    \begin{bmatrix}
        1 & 1 \\
        i & -i
    \end{bmatrix}
    \begin{bmatrix}
        \pa_{z_k} \\
        \pa_{z_k^*}
    \end{bmatrix} \hP \non\\
    &=& \int d^2 z_k e^{-|z_k|^2} \pa_{z_k}\hP. \label{eq:a8}
\end{eqnarray}
Substituting Eq.~(\ref{eq:a8}) into Eq.~(\ref{zt}), we obtain
\begin{eqnarray}
\mm(z_t^* \hP) &=& -i\sum_k g_k e^{i\om_kt} \int \frac{d^2 z}{\pi} e^{-|z|^2}\frac{\partial \hP}{\partial z_k} \non\\
&=& \fr{\pa z_t^*}{\pa z_k^*}\int \frac{d^2 z}{\pi} e^{-|z|^2}\frac{\partial \hP}{\partial z_k}.
\end{eqnarray}
Then, we apply the chain rule: 
\begin{eqnarray}
\frac{\partial (\cdot) }{\partial z_k} = \int_0^t ds \frac{\partial z_s}{\partial z_k} \frac{\delta (\cdot)}{\delta z_s},
\end{eqnarray}
and obtain the Novikov theorem\cite{novikov,QSD_Yu99},
\begin{eqnarray}
\mm(z_t^* \hP) &=& \fr{\pa z_t^*}{\pa z_k^*} \int \frac{d^2 z}{\pi} e^{-|z|^2} \int_0^t ds \frac{\partial z_s}{\partial z_k}\frac{\delta \hP}{\delta  z_s} \non \\
&=& \int \frac{d^2 z}{\pi} e^{-|z|^2} \int_0^t ds \Big[ \fr{\pa z_t^*}{\pa z_k^*} \fr{\pa z_s}{\pa z_k} \Big] \frac{\delta \hP}{\delta  z_s} \non \\
&=& \int_0^t ds \mm(z_t^*z_s) \mm(\frac{\delta \hP}{\delta  z_s}). 
\end{eqnarray}


\bibliography{positivity_meq}
\end{document}